\documentclass[letterpaper]{article} 
\usepackage{aaai2026}  
\usepackage{times}  
\usepackage{helvet}  
\usepackage{courier}  
\usepackage[hyphens]{url}  
\usepackage{graphicx} 
\usepackage{amsmath}
\usepackage{booktabs}
\usepackage{multirow}
\usepackage{natbib}  
\usepackage{caption} 
\usepackage{algorithm}
\usepackage{algorithmic}

\usepackage{newfloat}
\usepackage{listings}
\DeclareCaptionStyle{ruled}{labelfont=normalfont,labelsep=colon,strut=off} 
\floatstyle{ruled}
\newfloat{listing}{tb}{lst}{}
\floatname{listing}{Listing}
\title{Can We Predict the Next Question? A Collaborative Filtering Approach to Modeling User Behavior}
\author {
    Bokang Fu\textsuperscript{\rm 1},
    Jiahao Wang\textsuperscript{\rm 1},
    Xiaojing Liu\textsuperscript{\rm 1},
    Yuli Liu\textsuperscript{\rm 1}\thanks{*Corresponding author}
}
\affiliations {
    \textsuperscript{\rm 1}Qinghai University\\
    fubokang1123@gmail.com, wangjiahao2025@gmail.com, liuxj@qhu.edu.cn, liuyuli012@gmail.com
}

\usepackage{bibentry}

\begin{document}

\maketitle

\begin{abstract}

In recent years, large language models (LLMs) have excelled in language understanding and generation, powering advanced dialogue and recommendation systems. However, a significant limitation persists: these systems often model user preferences statically, failing to capture the dynamic and sequential nature of interactive behaviors. The sequence of a user's historical questions provides a rich, implicit signal of evolving interests and cognitive patterns, yet leveraging this temporal data for predictive tasks remains challenging due to the inherent disconnect between language modeling and behavioral sequence modeling.

To bridge this gap, we propose a Collaborative Filtering-enhanced Question Prediction (CFQP) framework. CFQP dynamically models evolving user-question interactions by integrating personalized memory modules with graph-based preference propagation. This dual mechanism allows the system to adaptively learn from user-specific histories while refining predictions through collaborative signals from similar users. Experimental results demonstrate that our approach effectively generates agents that mimic real-user questioning patterns, highlighting its potential for building proactive and adaptive dialogue systems.

\end{abstract}


\section{1  Introduction}

The 21st century has witnessed a paradigm shift in artificial intelligence, catalyzed by the advent of large language models (LLMs). Their profound ability to comprehend, generate, and reason with human language has unlocked transformative capabilities across a vast spectrum of applications, heralding a new era of language-centric AI. From powering sophisticated, human-like dialogue systems to enabling hyper-personalized recommendation engines, LLMs have become the cornerstone of modern human-computer interaction. They have demonstrated a remarkable capacity to process context, retain information over short conversational arcs, and produce fluent, relevant responses, thereby significantly enhancing user engagement and task success rates. As these models become increasingly integrated into our digital lives, the focus of research is naturally shifting from mere language proficiency to a deeper, more holistic understanding of the user, one that encompasses not just their explicit commands but also their implicit intentions, evolving interests, and behavioral patterns.

Despite their successes, the dominant paradigm in applying LLMs to interactive systems suffers from a critical and pervasive limitation: it is difficult to capture the dynamic and sequential nature of user preferences\cite{liu2022determinantal}. Current research predominantly operates on a static or, at best, a slowly adapting conception of the user. Systems are often designed to handle conversations as a series of independent or loosely connected turns, relying on user profiles that are either manually crafted or updated infrequently. This approach fundamentally misrepresents the reality of human engagement. A user's interests are not static entities, they are fluid, evolving constructs that shift within a single interaction session and across longer timescales. For instance, a user researching a topic may begin with broad queries and progressively narrow their focus, exhibiting a clear sequential dependency in their line of questioning. A static model is blind to this cognitive journey.

This issue points to a deeper, more fundamental chasm: the gap between language modeling and behavioral sequence modeling. While LLMs excel at predicting the next word in a sentence, they are not inherently designed to predict the user's next action or question in a sequence. A user's history of interaction is more than just a collection of utterances; it is a structured, time-ordered sequence that encodes their cognitive state, their information-seeking strategy, and their evolving intent.Ignoring this temporal structure resembles reading a book's pages in random order: while individual words remain intelligible, the coherent narrative is lost. Consequently, existing systems remain largely reactive. They can respond adeptly to what a user has just asked, but they lack the foresight to anticipate what the user might ask next. This reactive posture limits their ability to provide truly proactive, intelligent assistance, creating a ceiling on the quality and naturalness of the interaction.

The problem is twofold. First, conventional models often treat user history as a simple sequence of events, failing to distill a persistent, evolving model of individual user characteristics. Second, they fail to account for the valuable behavioral patterns found among similar user preferences\cite{liu2024learning}, a principle that forms the foundation of conventional recommendation systems. This gap presents a compelling research opportunity: to develop a framework that endows LLMs with both a long-term personal memory and a mechanism for collaborative intelligence, thereby transforming them from generic conversationalists into truly user-centric agents. Therefore, the central motivation of our work is to develop a system that can learn a user's dynamic behavioral patterns and leverage this understanding to predict their future actions.

Motivated by this vision, we introduce the Collaborative Filtering Question Prediction (CFQP) framework, a novel approach designed to synergistically integrate personalized memory with collaborative filtering to achieve superior user-specific question prediction. Our framework addresses the inherent limitations of current systems by constructing a dual-faceted user model. The core of our methodology lies in its ability to dynamically model user preferences by creating a personalized memory module for each user\cite{liu2024universal}. This module learns and adapts over time, capturing the unique evolution of a user's interests and interaction patterns, drawing inspiration from foundational personalization techniques like content-based and collaborative filtering. Simultaneously, CFQP implements a preference propagation mechanism across a user-user graph, allowing insights from like-minded users to inform and enrich the predictions for an individual. This collaborative dimension ensures that predictions are not only historically grounded but also socially informed, enabling the model to anticipate needs even when direct historical data is sparse.

The main contributions of this paper are summarized as follows:

\begin{itemize}
    \item A Novel Hybrid Framework: We propose CFQP, the first framework to our knowledge that combines a dedicated personalized memory module with a collaborative preference propagation network for the task of user question prediction. This hybrid architecture allows the model to capture both idiosyncratic user evolution and community-driven trends.
    \item Dynamic and Personalized User Modeling: We introduce a sophisticated memory module that creates a dynamic and persistent representation of each user. This goes beyond simple context-windowing to build a long-term understanding of user preferences and their temporal dynamics.
    \item Collaborative Intelligence Integration: We design and implement a preference propagation mechanism that leverages a user-user similarity. This allows the model to perform collaborative filtering in a novel context, enhancing prediction accuracy by sharing preference information among similar users.
\end{itemize}


\section{2 Related Work}

\textbf{Collaborative Filtering and User Preference Propagation}. Collaborative Filtering (CF), as one of the fundamental approaches in recommender systems, predicts target users' unknown preferences by leveraging known preferences from user communities\cite{zhao2023embeddingrecommendersystemssurvey}. However, traditional CF methods, such as user-based CF \cite{Su2009} and item-based CF, face significant challenges when dealing with variations in user rating behaviors and dynamic interactions with item attributes\cite{Vaghari2025}. The core issue lies in the static nature of these models, which often fail to capture the temporal evolution of user interests, a phenomenon known as concept drift \cite{Gama2014} . Recent studies demonstrate that user preferences are influenced not only by individual historical behaviors but also exhibit contextual variability. For instance, the same user may assign different importance weights to identical item attributes across different scenarios\cite{Vaghari2025}. To address these limitations, researchers have explored integrating more complex models, such as graph neural networks (GNNs), to model the intricate user-item relationships \cite{Wei2019} and autoencoders for learning latent representations \cite{Sedhain2015} . Furthermore, the rise of Large Language Models (LLMs) has opened new avenues for enhancing CF by distilling world knowledge and reasoning capabilities into the recommendation process, thereby better addressing issues like data sparsity and cold starts \cite{sun2024largelanguagemodelsenhanced} .

\noindent\textbf{Personalized Large Language Models}. Large Language Models (LLMs) have demonstrated remarkable capabilities in natural language processing, yet they still face significant challenges when handling user-specific personalized tasks, such as comprehending individual emotions, writing styles, and preferences\cite{liu2025surveypersonalizedlargelanguage}. The goal of personalization is to bridge the gap between generic models and individual user needs, transforming passive information filtering into active user engagement \cite{Chen_2024} . To address this limitation, Personalized Large Language Models (PLLMs) leverage user-specific data (including user profiles, historical dialogues, content, and interaction records) to deliver tailored responses. From a technical perspective, current research on PLLMs primarily focuses on three key dimensions: personalized prompting at the input level, personalized adapters at the model architecture level, and personalized alignment at the objective level\cite{zhang2025personalizationlargelanguagemodels}. These approaches aim to create models that can not only understand a user's explicit requests but also infer their implicit intents and preferences from long-term interactions, a critical step towards more human-like conversation \cite{woźniak2024personalizedlargelanguagemodels} .

\noindent\textbf{Asking Clarifying Questions}. Current research on Asking Clarifying Questions (ACQs) remains predominantly confined to passive responses to immediate ambiguous queries, lacking dynamic modeling capabilities for user preferences and knowledge states\cite{Tavakoli2020}. Existing systems typically determine whether to ask questions based solely on the current conversational turn, failing to predict users' latent clarification needs or proactively anticipate follow-up questions \cite{rahmani2023surveyaskingclarificationquestions}. This reactive nature limits their effectiveness in complex, multi-turn information-seeking dialogues \cite{Radlinski2017} . Although some studies have established ACQ generation frameworks in open-domain dialogues\cite{Aliannejadi_2019}, these single-agent paradigms still fail to break through the limitations of reactive clarification. Recent work has begun to explore more proactive strategies, such as using negative feedback to refine subsequent questions \cite{Bi2021} or developing models that can strategically drive the conversation forward \cite{Zhu_2021}. However, these approaches generally neglect the critical value of mining collaborative signals from other users' interaction patterns, a gap that our proposed framework aims to fill by integrating collaborative intelligence into the proactive question-asking process.

\section{3 Proposed Model:CFQP}

\begin{figure*}[h]
	\centering
	\includegraphics[width=0.95\linewidth]{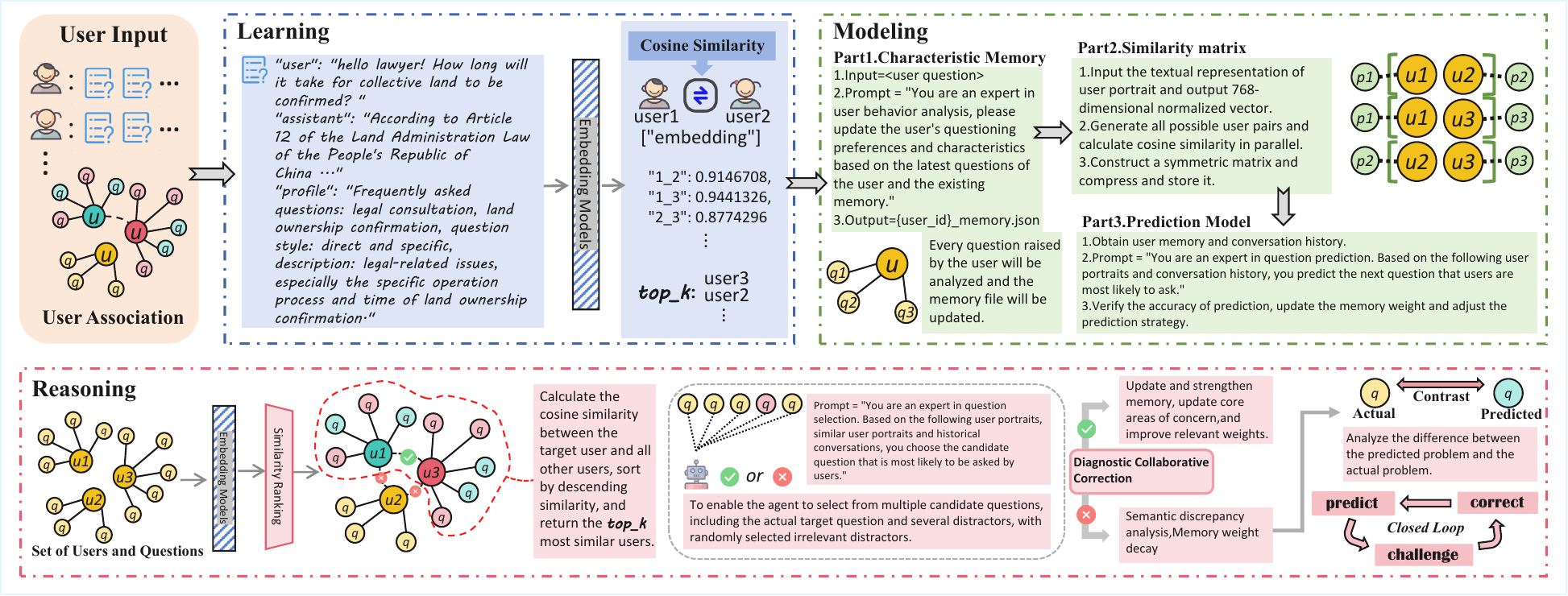}
        \caption{The overall architecture of the CFQP framework.The whole process is coordinated by three core modules: Learning, Modeling and Reasoning.}
	\label{fig:overall}
\end{figure*}

\subsection{3.1 Overall Framework}

To achieve accurate prediction of users' sequential questioning behavior, we propose an innovative Collaborative Filtering enhanced Question Prediction (CFQP) framework. The core idea of this framework is to predict a target user's next question by integrating their personalized dynamic preferences with the collective intelligence of similar user groups. As illustrated in Figure 1, the CFQP framework primarily consists of three collaboratively functioning modules: Learning, Modeling, and Reasoning.

\subsection{3.2 User Preference Learning and Collaborative Signal Mining}

The initial step of the CFQP framework is to construct a rich and semantically meaningful representation for each user in the system. This process aims not only to capture the static profile of individual users but, more critically, to unearth latent collaborative signals among them, laying the groundwork for subsequent collaborative filtering and reasoning. 

\noindent\textbf{Comprehensive User Profiling}. We begin by aggregating all available textual information for each user \(u_i\) to build a comprehensive descriptive document, \(D_i\). As illustrated in the Learning module of Figure 1, this document \(D_i\) is dynamically composed of two parts: 

\begin{itemize}

\item Static Profile: Explicitly provided personal information from the user, such as professional background, interest tags, or a self-description within the system. This information provides a baseline for the user's long-term, stable interests. 

\item Dynamic Interaction History: The complete dialogue records between the user and the system, including all questions posed by the user and the corresponding answers provided by the system. This data reflects the user's immediate interests, knowledge exploration paths, and the evolution of their cognitive patterns over specific periods.

\end{itemize}

By integrating these two components, the document \(D_i\) comprehensively represents a user, encompassing both their invariant core characteristics and their dynamically evolving query intentions.

\noindent\textbf{Embedding-based User Representation}. To enable the model to understand the unstructured text data \(D_i\), we employ a powerful pre-trained language model, specifically BGE (BAAI/bge-base-zh), to map it into a dense vector space. We chose BGE for its superior performance in generating sentence- and paragraph-level semantic representations, especially optimized for Chinese text, ensuring that semantically similar texts are closer in the vector space.

For each user's document \(D_i\), the BGE model generates a fixed-dimensional embedding vector \(\mathbf{v}_i \in \mathbf{R}^d\) (e.g., \(d=768\) in our implementation). This vector \(\mathbf{v}_i\) becomes the digital representation of user \(u_i\), encoding not only the topics they are interested in but also implicitly their questioning style, cognitive level, and potential knowledge gaps. The use of BGE ensures high-quality embeddings tailored to Chinese semantic understanding tasks, such as retrieval and clustering, while maintaining computational efficiency.

\noindent\textbf{User Similarity Calculation}. The core principle of collaborative filtering lies in the assumption that "similar people will exhibit similar behaviors." To quantify the similarity between users, we calculate the cosine similarity between the representation vectors of every pair of users \((u_i, u_j)\):

\[
\text{sim}(u_i, u_j) = \frac{\mathbf{v}_i \cdot \mathbf{v}_j}{\|\mathbf{v}_i\| \|\mathbf{v}_j\|}
\]

This similarity score, ranging from -1 to 1, measures the closeness of two users in terms of their interests and cognitive patterns. We perform this calculation for all user pairs in the system, thereby constructing an \(N \times N\) symmetric user similarity matrix \(\mathbf{S}\), where \(N\) is the total number of users and \(\mathbf{S}_{ij} = \text{sim}(u_i, u_j)\).

This matrix \(\mathbf{S}\) can be viewed as a weighted undirected graph where nodes are users and edge weights represent their similarity. It forms the basis for the "User Association" depicted in Figure 1 and serves as the bridge for preference propagation among users.

Based on the constructed similarity matrix \(\mathbf{S}\), for any target user \(u_t\), we can easily find the top-k most similar users by sorting the values in the corresponding row (or column) of the matrix \(\mathbf{S}\). We denote this set of users as \(N_k(u_t)\).

This "neighborhood" set \(N_k(u_t)\) will inject the historical behaviors and preferences of similar users into the prediction process of the next question of the target user \(u_t\) in the subsequent reasoning module, effectively mitigating data sparsity and cold-start problems that may arise when relying solely on individual historical data.

\subsection{3.3 Dynamic Preference Modeling}

A user's interests are not static; they continuously evolve through interaction with the system. To accurately capture this time-varying nature, we designed the Dynamic Preference Modeling module. This module acts as a bridge connecting a user's historical behavior with their future intent, integrating personalized memory, collaborative signals, and the reasoning capabilities of large language models to construct an adaptively updated user model.

\noindent\textbf{Characteristic Memory}. To track the user's interest drift in real-time, we maintain a lightweight yet informative dynamic memory file, \(M_i\), for each user \(u_i\). This memory file (a JSON object in our implementation) stores the core topics and concepts the user is focused on, along with their corresponding weights, forming a weighted preference snapshot. Whenever a user poses a new question, \(q_{new}\), a specialized LLM agent is activated. Its task is to parse the semantics of the new question and update the existing memory \(M_i\). This update process may involve increasing the weight of topics related to \(q_{new}\), introducing new topics, or decaying the weights of older ones. This mechanism enables the user model to sensitively reflect the latest points of interest, providing immediate context for prediction.

\noindent\textbf{Similarity Matrix}. While user preferences are dynamic, the underlying similarity between users is relatively stable in the short term. The user similarity matrix \(\mathbf{S}\), calculated in the first stage, serves as a static structure in this module, defining the collaborative relationship network among users. This matrix acts as the backbone for preference propagation and collaborative reasoning, allowing the system to borrow information for a target user from the behavior of similar users.

\noindent\textbf{Prediction Model}. This is the core engine for generating the final prediction. We again leverage the contextual understanding and generation capabilities of an LLM to construct a sophisticated prediction prompt. This prompt is carefully designed to fuse information from various sources and guide the LLM toward high-quality reasoning. Specifically, the context provided to the prediction model includes: (1) The target user's personal information: their dynamically updated characteristic memory \(M_t\) and recent conversation history \(H_t\); (2) Collaborative signals: profile summaries and typical questions from their top-k similar users \(N_k(u_t)\). By integrating this personalized, dynamic, and collaborative information, we prompt the LLM to predict the next question, \(\hat{q}_{t+1}\), that the target user \(u_t\) is most likely to ask. 

\begin{table*}[t]
\small
  \centering
    \begin{tabular}{l|ccc|ccc} 
    \toprule
     \multirow{2.5}{*}{Method} 
     & \multicolumn{3}{c}{LexRAG} 
     & \multicolumn{3}{c}{CrossWOZ} \\  
     \cmidrule(lr){2-4} \cmidrule(lr){5-7}  
     & Jaccard & Cosine Similarity & LLM as a Judge  & Jaccard & Cosine Similarity & LLM as a Judge \\  
    \midrule
    Qwen  & 0.1173 & 0.8061 & 0.2440 & 0.0828 & 0.7195 & 0.2224\\
    Deepseek   & 0.0974 & 0.7867 & 0.2704 & 0.0945 & 0.7290 & 0.2192\\
    GLM   & 0.1233 & 0.7955 & 0.2630 & 0.1133 & 0.7269 & 0.2030\\
    CFQP & \textbf{0.1810} & \textbf{0.8249} & \textbf{0.4500} & \textbf{0.1350} & \textbf{0.7653} & \textbf{0.2627}\\
    \bottomrule
    \end{tabular}%
    \caption{The CFQP framework is compared with three baseline models. The best method has been bolded.}
\end{table*}

\subsection{3.4 Collaborative Reasoning and Error Correction}

The Collaborative Reasoning and Error Correction module is key to the CFQP framework's ability to achieve continuous self-optimization and ensure that model predictions are precisely aligned with the user's true preferences. This module not only evaluates the accuracy of predictions but, more importantly, dynamically adjusts the model's internal state (user memory) using collaborative signals, thereby creating a closed loop for diagnostic error correction.

\noindent\textbf{Collaborative Reasoning and Prediction}. Upon receiving the user's current query \( q_t \), the model first generates a prediction for the next question, \( \hat{q}_{t+1} \), based on the output from the Dynamic Preference Modeling module. This prediction integrates the user's own historical interaction sequence and preference information obtained from similar neighbors through the collaborative network.

\[ \hat{q}_{t+1} = \text{LLM}(M_u, N_u, q_t) \]

Where \( M_u \) is the memory of user \( u \), and \( N_u \) is the set of their similar neighbors.

\noindent\textbf{Candidate Question Selection}. To effectively evaluate and correct the model, we designed a challenge mechanism rather than simply comparing the prediction with the actual next question \( q_{t+1} \). Specifically, we construct a candidate question set \( C_q \) that includes:

\begin{itemize}

\item Ground Truth: The actual question \( q_{t+1} \) that the user asks in the next turn.

\item Distractors: A set of questions randomly sampled from the question corpus, which may be contextually relevant or entirely irrelevant to the user's current session. These distractors are designed to challenge the model's ability to distinguish true user intent from noise.

\end{itemize}

The model is then tasked with selecting the most likely next question from \( C_q \).

\[ q^{*}_{t+1} = \text{argmax}_{q' \in C_q} P(q' | M_u, N_u, q_t) \]

Here, \( P(q' | M_u, N_u, q_t) \) is the probability that the model predicts the user will ask question \( q' \) in the current state.

\noindent\textbf{Diagnostic Collaborative Correction}. The correction mechanism is triggered based on the model's performance in the selection task. 

\begin{itemize}

\item If the model chooses correctly (i.e., \( q^{*}_{t+1} = q_{t+1} \)): This indicates that the model's current internal state \( M_u \) accurately reflects the user's preferences. Therefore, we reinforce the memory components related to the current query \( q_t \) and the prediction \( q_{t+1} \) through a positive reinforcement operation.
    \[ M_u \leftarrow \text{Reinforce}(M_u, q_t, q_{t+1}) \]

\item If the model chooses incorrectly (i.e., \( q^{*}_{t+1} \neq q_{t+1} \)): This exposes a bias in the model's understanding. The incorrect choice points to a "knowledge gap" or "preference misjudgment" within the model's internal state. At this point, the error correction mechanism is activated: (1) Modal Analysis: Analyze the discrepancy between the incorrect choice \( q^{*}_{t+1} \) and the ground truth \( q_{t+1} \), then analyze the difference between the predicted problem \( \hat{q}_{t+1} \) and the actual problem \( q_{t+1} \). (2) Perform Attenuation/Correction: Correct the memory or cooperative signal that leads to wrong prediction, reduce the memory weight of wrong selection, and keep the error record for subsequent analysis. 

\end{itemize}

Through this "predict-challenge-correct" closed loop, the CFQP model can continuously learn from its mistakes and dynamically adjust its understanding of user preferences, thereby providing increasingly accurate next-question predictions in successive interactions.

\section{4 Experiments}

In this section, we conduct experiments to evaluate our proposed approach, and analyze the empirical results.

\subsection{4.1 Experimental Setup}

\noindent\textbf{Datasets}. To comprehensively evaluate the effectiveness of our proposed CFQP framework in different scenarios, we employ two representative datasets: LexRAG and CrossWOZ . LexRAG\cite{li2025lexragbenchmarkingretrievalaugmentedgeneration} is a specialized English legal consultation benchmark for testing the model's deep reasoning and knowledge retrieval capabilities in a specific domain. CrossWOZ\cite{zhu2020crosswozlargescalechinesecrossdomain} is a large-scale Chinese cross-domain task-oriented dialogue dataset for examining the model's adaptability and robustness in complex dialogue scenarios involving multiple topics and intents. By conducting experiments on these two datasets, we can thoroughly assess CFQP's comprehensive performance in handling both single-domain deep dialogues and cross-domain task transitions. 

\noindent\textbf{Implementation Details}. Our CFQP framework is implemented based on PyTorch. The user representation vector is generated by the BGE (BAAI/bge-base-zh) model. In the comparison and replacement experiment, we integrated three mainstream large-scale language models as the backbone: Qwen, Deepseek, and GLM. Unless otherwise specified, the default CFQP model uses GLM as its backbone LLM. 

\noindent\textbf{Evaluation Metrics}. We use the following indicators to evaluate the model performance: (1) Jaccard Similarity: By calculating the ratio of the intersection and union of word sets between the prediction problem and the real problem, the degree of overlap at the lexical level is measured. This index is sensitive to the matching degree of keywords. (2) Cosine Similarity: The prediction problem and the real problem are transformed into sentence vectors by using BGE, and the similarity at the semantic level is evaluated by calculating the cosine values between the vectors. (3) LLM as a referee (LLM-as-a-Judge): Invite a powerful third party LLM (such as GLM) to grade the prediction quality (0-1 points), and the evaluation dimensions include relevance, fluency and logicality, so as to provide evaluation results closer to human judgment. (4) Accuracy of Question Selection: It is used to evaluate the model's ability to select correct questions from a candidate set containing real answers and interference items in the "diagnostic collaborative error correction" link. 

\subsection{4.2 Baselines}

In order to verify the superiority of CFQP framework, we selected the following three powerful large-scale language models as the baseline. When testing, these models will receive the same user conversation history as CFQP as In-context Learning, and be asked to directly predict the user's next question. (1) Qwen: A universal and excellent large-scale language model. (2) Deepseek: A large language model with expertise in code and logical reasoning. (3) GLM: A universal dialogue model with bilingual ability.

\begin{table*}[t]
\scriptsize
  \centering
    \begin{tabular}{l|cccc|cccc} 
    \toprule
     \multirow{2.5}{*}{Method} 
     & \multicolumn{4}{c}{LexRAG} 
     & \multicolumn{4}{c}{CrossWOZ} \\  
     \cmidrule(lr){2-5} \cmidrule(lr){6-9}  
     & Jaccard & Cosine Similarity & LLM as a Judge & Accuracy (\%) & Jaccard & Cosine Similarity & LLM as a Judge & Accuracy (\%)\\  
    \midrule
    CFQP-Spark & 0.1735 & 0.8172 & 0.4170 & 85.00 & \textbf{0.1366} & 0.7649 & 0.2580 & 55.00\\
    CFQP-Hunyuan & 0.1566 & 0.8194 & 0.3920 & 90.00 & 0.1270 & 0.7514 & 0.2553 & \textbf{56.67}\\
    CFQP-GLM & \textbf{0.1810} & \textbf{0.8249} & \textbf{0.4500} & \textbf{93.33} & 0.1350 & \textbf{0.7653} & \textbf{0.2627} & 50.00\\
    \bottomrule
    \end{tabular}%
    \caption{Comparison of different backbone models in CFQP framework. The best method has been bolded.}
\end{table*}

\subsection{4.3 Performance Comparison with Standalone Large Models}

\noindent\textbf{Objective}. The core goal of this experiment is to verify whether our proposed CFQP framework has a significant performance advantage in predicting the user's next problem task compared with the current mainstream independent large-scale language model that only relies on In-context Learning. At present, the standard paradigm of directly using LLM to process serialization tasks is to take its dialogue history as long context input. However, this method has fundamental defects: it lacks the structural modeling ability of users' long-term and dynamic preferences, and it cannot use group intelligence for collaborative reasoning. We aim to prove through this experiment that CFQP framework can understand users' intentions more deeply and make more accurate predictions through its personalized memory module and cooperative signal mechanism among users.

\noindent\textbf{Methodology}. We directly confront the CFQP framework (the default backbone model is GLM) with three powerful baseline models (Qwen, Deepseek, GLM). In order to ensure fairness, all models will receive exactly the same user history dialogue as input when facing the same prediction task. The baseline model directly takes these histories as context cues, while the CFQP framework will make comprehensive reasoning on this basis.

We will fully run this experiment on two data sets, LexRAG and CrossWOZ, and use three indicators, namely Jaccard similarity, cosine similarity and llm as a judge score, to evaluate the prediction output of all models. Jaccard similarity is evaluated from lexical level, cosine similarity is evaluated from semantic level, and llm as a judge score provides a quality evaluation closer to human comprehensive judgment.

\noindent\textbf{Results and Analysis}. The experimental results, as shown in Table 1, clearly show that our proposed CFQP framework is consistent and significantly superior to all baseline models in all test scenarios. On both LexRAG and CrossWOZ datasets, CFQP has achieved the best performance on three key indicators: Jaccard similarity, cosine similarity and LLM as a judge score. This overwhelming advantage strongly confirms our core hypothesis: the traditional LLM paradigm, which only relies on long-term context for learning, has inherent defects in understanding users' long-term and dynamic intentions, and CFQP framework can effectively make up for this deficiency through its innovative structured design.

Personalized memory module refines the flat dialogue history into a structured user preference model, and realizes the transformation from "passive memory" to "active understanding". This is directly reflected in the leading similarity between Jaccard and cosine, which means that CFQP can capture the key words and core semantics of users' next demand more accurately. This module is particularly effective when dealing with information-intensive single-domain dialogues such as LexRAG, and can continuously track the user's intention trajectory of in-depth exploration. The collaborative signal module introduces "group intelligence" to assist reasoning by learning from the behavior patterns of similar users. When the user's history is insufficient, the collaborative signal provides a powerful supplement, making the prediction more forward-looking.

To sum up, the experimental results strongly prove that the CFQP framework can understand the user's intention more deeply than the independent large model that only relies on context learning by organically combining personalized memory and collaborative signals, thus making a substantial performance breakthrough in predicting the user's next problem.

\begin{table*}[t]
\scriptsize
  \centering
    \begin{tabular}{l|cccc|cccc} 
    \toprule
     \multirow{2.5}{*}{Method} 
     & \multicolumn{4}{c}{LexRAG} 
     & \multicolumn{4}{c}{CrossWOZ} \\  
     \cmidrule(lr){2-5} \cmidrule(lr){6-9}  
     & Jaccard & Cosine Similarity & LLM as a Judge & Accuracy (\%) & Jaccard & Cosine Similarity & LLM as a Judge & Accuracy (\%)\\  
    \midrule
    CFQP-NoC  & 0.1093 & 0.8092 & 0.3600 & 68.00 & 0.0892 & 0.7400 & 0.2350 & 28.00\\
    CFQP-NoM   & 0.1191 & 0.8053 & 0.2700 & 66.00 & 0.0970 & 0.7314 & 0.2510 & 34.00\\
    CFQP-NoS   & 0.1224 & 0.8033 & 0.3200 & - & 0.0810 & 0.7216 & 0.2390 & -\\
    CFQP & \textbf{0.1810} & \textbf{0.8249} & \textbf{0.4500} & \textbf{93.33} & \textbf{0.1350} & \textbf{0.7653} & \textbf{0.2627} & \textbf{50.00}\\
    \bottomrule
    \end{tabular}%
    \caption{The CFQP framework is compared with three three ablation variants. The best method has been bolded.}
\end{table*}

\subsection{4.4 Comparison of Different Backbone Models within the CFQP Framework}

\noindent\textbf{Objective}. The purpose of this experiment is to explore the universality, robustness and expansibility of CFQP framework. We want to answer two questions: First, can CFQP framework be effectively integrated with large-scale language models with different capabilities and different architectures? Second, as the backbone of the framework "reasoning core", how does LLM's own ability affect the final performance of CFQP? Through this experiment, we intend to prove that CFQP is a flexible and powerful "enhancer", which can effectively amplify and utilize the specific advantages of different LLM.

\noindent\textbf{Methodology}. We replaced LLM, the backbone of CFQP framework, with Spark, Hunyuan and GLM respectively, thus constructing three model variants with complete functions but different "brains": CFQP-Spark, CFQP-Hunyuan and CFQP-GLM. These three model variants will complete the next problem prediction task on two data sets under the same experimental conditions.

\noindent\textbf{Results and Analysis}. Experiments have successfully verified the high flexibility and robustness of CFQP framework as a general "enhancer", as shown in Table 2. This framework can integrate and effectively utilize LLM with different architectures and capabilities. (1) Performance is strongly related to the core LLM: the final performance of CFQP framework is directly proportional to the ability of its built-in LLM "brain". It is proved that the framework can effectively enlarge the advantages of the backbone model. (2) Universal validity: All model variants scored more than the single large model in the next problem prediction task, which confirmed the universal applicability and effectiveness of the CFQP framework. 

In a word, CFQP is a powerful framework with high flexibility and robustness, which allows developers to flexibly choose the best LLM core according to their needs to achieve the best application effect.

\subsection{4.5 Ablation Study}

\noindent\textbf{Objective}. The main purpose of this ablation study is to analyze the proposed CFQP framework strictly and quantitatively to verify the specific contribution of each core architecture component. Although the whole framework is designed as an integrated system, the key is to prove that its overall performance is not only the result of a single leading module, but the result of the synergy of all its parts.

\noindent\textbf{Methodology}. Based on the default CFQP model, we designed the following three ablation variants: (1) CFQP w/o Collaborative Signal (CFQP-NoC): Remove the inter-user preference propagation module. The model no longer receives collaborative signals from similar users \( N_u \) in prediction, but only relies on users' personalized memory and conversation history. (2) CFQP w/o Memory (CFQP-NoM): Remove the personalized memory module \( M_u \). The model can't access the user's long-term and dynamic preference snapshot, and only relies on the real-time conversation history and collaborative signals. (3) CFQP w/o Selection (CFQP-NoS): Remove the problem selection and feedback mechanism in the diagnostic collaborative error correction module. The model directly generates predictions, and its internal memory \( M_u \) is no longer optimized through the "challenge-error correction" cycle.

\noindent\textbf{Results and Analysis}. Ablation experiment strongly proves the indispensability and synergistic effect of three core modules in CFQP framework, as shown in Table 3. Removing any module will lead to a significant decline in model performance, which verifies the integrity and efficiency of the framework design. (1) Collaborative Signal: it is the key to tap the collective wisdom. Without it, it is difficult for the model to go beyond the limitations of individual users' historical data and capture complete potential interests. (2) Personalized Memory is the cornerstone of deep personalization. Without it, the model will become "short-sighted" and cannot understand the continuity and dynamic evolution of users' cognitive paths. (3) Diagnostic Selection: it is a mechanism to ensure continuous self-optimization. Without it, the model will lose the ability to correct deviation and continuously align users' real preferences.

In a word, these three modules complement each other and jointly build a powerful forecasting system that can not only make use of collective wisdom, but also realize deep personalization and adaptive learning.

\section{5 Conclusion}

This paper aims to solve the problems of insufficient personalization and inaccurate capture of users' intentions when the existing large-scale language models generate follow-up questions in multiple rounds of dialogue. In order to meet this challenge, we innovatively propose a personalized follow-up problem prediction framework CFQP based on collaborative filtering. By integrating users' own historical interaction information with similar users' behavior patterns, the framework significantly improves the personalization level and relevance of subsequent question generation.

The principles underlying the CFQP framework have significant implications for the broader field of Personalized Large Language Models (PLLMs). Current PLLMs excel at passively adapting to a user's historical data, such as writing style or known preferences. Our work introduces a paradigm shift from this reactive personalization towards proactive, predictive personalization . The hybrid approach of CFQP integrates dynamic personal memory with collaborative intelligence from the user community, providing a specific architectural blueprint for the next-generation pllm. By integrating such a framework, future PLLMs could evolve from being merely responsive assistants into truly proactive companions that can anticipate user needs, foresee information gaps, and guide conversations more intelligently and empathetically. This represents a crucial step towards building more autonomous and user-centric AI systems. 

Although the CFQP framework has achieved encouraging results, there are still some limitations in this study, leaving room for future work: 
(1)Sparsity of collaborative filtering and cold start problem: for new users or users with sparse interaction, the effectiveness of collaborative signals may be reduced. 
(2)Optimization of user memory module: We can further study more complex memory network structure, such as introducing forgetting mechanism to better capture the dynamic changes of user interests. 
(3)Generalization ability of the framework: This research is mainly verified on a specific data set. In the future, CFQP framework needs to be tested and optimized in more diversified and broader fields (such as e-commerce, education, psychological consultation, etc.) to test its generalization ability. 
We believe that through in-depth exploration in the above direction, personalized follow-up question generation technology will better serve users and build a more intelligent, natural and humanized dialogue system.

\bibliography{aaai2026}


\end{document}